\documentclass[12pt]{iopart}

\usepackage{graphicx}

\usepackage{iopams}

\newtheorem{theorem}{Theorem}

\newtheorem{corollary}{Corollary}

\newtheorem{example}{Example}

\newtheorem{proposition}{Proposition}
\newtheorem{remark}{Remark}

\newenvironment{proof}[1][Proof]{\textbf{#1.} }{\ \rule{0.5em}{0.5em}}

\begin{document}

\title[]{Solvable Lie algebras with naturally graded
nilradicals and their invariants}

\author{J M Ancochea\dag, R. Campoamor-Stursberg\dag and L Garcia Vergnolle\dag}

\address{\dag\ Dpto. Geometr\'{\i}a y Topolog\'{\i}a\\Fac. CC. Matem\'aticas\\Universidad Complutense de Madrid\\Plaza de Ciencias, 3\\E-28040 Madrid, Spain}

\ead{ancochea@mat.ucm.es, rutwig@mat.ucm.es, lucigarcia@mat.ucm.es}

\begin{abstract}
The indecomposable solvable Lie algebras with graded nilradical of maximal nilindex and a Heisenberg subalgebra of codimension one are analyzed, and their generalized Casimir invariants calculated. It is shown that rank one solvable algebras have a contact form, which implies the existence of an associated dynamical system. Moreover, due to the structure of the quadratic Casimir operator of the nilradical, these algebras contain a maximal non-abelian quasi-classical Lie algebra of dimension $2n-1$, indicating that gauge theories (with ghosts) are possible on these subalgebras.

\end{abstract}


\maketitle



\section{Introduction}

Lie algebras and their invariants play a relevant role in physical models, like the multidimensional cosmological models, the Standard model, nuclear collective motions and  rotational states in particle and nuclear physics, the Petrov classification in General Relativity, quantum mechanics or, more recently, string theory, where isometry groups of high dimensional spaces are needed \cite{Ba,Lya,Pet}. Lie algebras also appear as symmetries of dynamical systems, and are therefore deeply related to the conservation laws of physics \cite{Sm}. Although semisimple Lie algebras occupy a central position whitin the Lie algebras appearing in physical models (Lorentz algebra, $\frak{su}(N)$ and $\frak{so}(p,q)$ series, symplectic algebras $\frak{sp}(N,\mathbb{R})$, etc.), as well as various semidirect products, like the Poincar\'e or the inhomogeneous Lie algebras, the class of solvable algebras has shown to be of considerable interest, as follows from their applicability to the theory of completely integrable Hamiltonian systems or non-abelian gauge theories \cite{Ar,Ok2}. While the classification of semisimple Lie algebras constitutes a classical result, solvable Lie algebras over the real field $\mathbb{R}$ have been classified only up to dimension six (\cite{Mo,Mu63,Tu} and references therein). The absence of global structural properties for this class, as well as the existence of parametrized families, indicate that a global classification is not feasible. However, proceeding by structure, important classes have been classified and their invariants determined, like solvable Lie algebras with Heisenberg or triangular nilradical \cite{Wi1,Wi2}, or specific types of solvable Lie algebras, called rigid ones \cite{Ca3}.

In \cite{Wi} the authors analyzed solvable Lie algebras having as nilradical a nilpotent Lie algebra $\frak{n}_{n,1}$ of maximal nilindex and an abelian ideal of codimension one. These algebras are interesting in the sense that the nilradical is naturally graded and has the maximal possible nilindex for a nilpotent Lie algebra. Indeed any nilpotent Lie algebra of dimension $n$ and nilindex $n-1$ can be shown to be a deformation of $\frak{n}_{n,1}$. Moreover, it is the only naturally graded algebra in odd dimension with this properties. In even dimension there exists another naturally graded algebra with maximal nilindex, called $Q_{2n}$, which contains a maximal Heisenberg subalgebra of codimension one. The purpose of this work is to analyze the real solvable Lie algebras having the latter nilradical, thus completing the study of invariants of solvable algebras having naturally graded nilradicals of maximal nilindex. While various properties like the structure of the generalized Casimir invariants are similar to the case studied in \cite{Wi}, the algebras of this paper present some particularities that make them worthy to be analyzed in detail, and that correspond to properties they are lost by contractions. Among these properties there is the constant number of invariants for any dimension, as well as the existence of linear contact forms, which allow the construction of a dynamical system on the corresponding groups and find applications in classical Hamiltonian structures. We also show that the $Q_{2n}$ algebras contain non-abelian quasi-classical nilpotent Lie algebras of codimension one, which are of certain interest for the study of solutions of the Yang-Baxter equations and in gauge theories based on solvable algebras \cite{Ok3}.

\medskip\ We apply the Einstein convention and usual notations for tensor algebra. By indecomposable Lie algebras we mean algebras that do not split into a direct sum of ideals. Unless otherwise stated, any Lie algebra considered in this work is defined over the fields $\mathbb{F}=\mathbb{R},\mathbb{C}$.

\section{Invariants of Lie algebras}

The invariant operators of the coadjoint representation of Lie algebras provide important information on a physical system, like quantum numbers, energy spectra or the existence of invariant forms. Polynomial invariants are traditionally called Casimir invariants, and occur for semisimple and nilpotent Lie  algebras. More generally, algebraic Lie algebras always admit invariants that are rational. For non-algebraic Lie algebras, specially for those that are solvable, we find rational or even transcendental invariants. These also find applications in representation theory or in classical integrable Hamiltonian systems \cite{Ar,Tro}. In fact, the algorithm usually applied to compute these invariants \cite{Be,Pe}, based on a system of linear first order partial differential equations, does not exclude the existence of irrational invariants, nor there is any physical reason for the invariants to be polynomials. In analogy with the classical Casimir operators, nonpolynomial invariants are called generalized Casimir invariants.

Let $\left\{  X_{1},..,X_{n}\right\}  $ be a
basis of $\frak{g}$ and $\left\{  C_{ij}^{k}\right\}  $ be the structure
constants over this basis. We consider the representation of $\frak{g}$ in the
space $C^{\infty}\left(  \frak{g}^{\ast}\right)  $ given by:
\begin{eqnarray}
\widehat{X}_{i}=-C_{ij}^{k}x_{k}\partial_{x_{j}},
\end{eqnarray}
where $\left[  X_{i},X_{j}\right]  =C_{ij}^{k}X_{k}$ $\left(  1\leq i<j\leq
n,\;1\leq k\leq n\right)  $. This representation is easily seen to satisfy the
brackets $\left[  \widehat{X}_{i},\widehat{X}_{j}\right]  =C_{ij}^{k}%
\widehat{X}_{k}$. The invariants are functions on the generators $F\left(
X_{1},..,X_{n}\right)  $ of $\frak{g}$ such that:
\begin{eqnarray}
\left[  X_{i},F\left(  X_{1},..,X_{n}\right)  \right]  =0,
\end{eqnarray}
and are found by solving the system of linear first order partial differential
equations:
\begin{eqnarray}
\widehat{X}_{i}F\left(  x_{1},..,x_{n}\right)  =-C_{ij}^{k}x_{k}%
\partial_{x_{j}}F\left(  x_{1},..,x_{n}\right)  =0,\;1\leq i\leq n. \label{sys}
\end{eqnarray}
and then replacing the variables $x_{i}$ by the corresponding generator
$X_{i}$ (possibly after symmetrizing). A maximal set of functionally
independent solutions of (\ref{sys}) will be called a fundamental
set of invariants. The cardinal $\mathcal{N}\left(  \frak{g}\right)  $ of such
a set can be described in terms of the dimension and a certain matrix
associated to the commutator table. More specifically, denote by $A\left(
\frak{g}\right)  $ the matrix representing the commutator table of $\frak{g}$
over a given basis, i.e.,
\begin{eqnarray}
A(\frak{g})=\left(  C_{ij}^{k}x_{k}\right)  .
\end{eqnarray}
Such a matrix has necessarily even rank by antisymmetry. Then $\mathcal{N}%
\left(  \frak{g}\right)  $ is given by
\begin{eqnarray}
\mathcal{N}\left(  \frak{g}\right)  =\dim\,\frak{g}-\mathrm{rank}\,\left(
C_{ij}^{k}x_{k}\right)  . \label{BB}
\end{eqnarray}
This formula was first described by Beltrametti and Blasi \cite{Be}. With
respect to the number of independent Casimir operators of $\frak{g}$, formula
(\ref{BB}) is merely an upper bound. For high dimensional Lie algebras, it is sometimes convenient to work with the analogue of formula (\ref{BB}) in terms of differential forms. Let $\mathcal{L}(\frak{g})=\mathbb{R}\left\{  d\omega_{i}\right\}  _{1\leq i\leq\dim\frak{g}}$ be the linear subspace of $\bigwedge^{2}\frak{g}^{\ast}$
generated by the Maurer-Cartan forms $d\omega_{i}$ of $\frak{g}$. If $\omega=a^{i}d\omega
_{i}\,\;\left(  a^{i}\in\mathbb{R}\right)  $ is a generic  element of $\mathcal{L}%
(\frak{g})$, there always exists an integer $j_{0}\left(  \omega\right)  \in\mathbb{N}$ such
that
\begin{eqnarray}
\bigwedge^{j_{0}\left(  \omega\right)  }\omega\neq0,\quad\bigwedge
^{j_{0}\left(  \omega\right)  +1}\omega\equiv0.
\end{eqnarray}
This equation shows that $r\left(  \omega\right)  =2j_{0}\left(
\omega\right)  $ is the rank of the 2-form $\omega$. We now define
\begin{eqnarray}
j_{0}\left(  \frak{g}\right)  =\max\left\{  j_{0}\left(  \omega\right)
\;|\;\omega\in\mathcal{L}(\frak{g})\right\}  .
\end{eqnarray}
The quantity $j_{0}\left(  \frak{g}\right)  $, which depends only on the
structure of $\frak{g}$, constitutes a numerical invariant of $\frak{g}$ \cite{C43}. The number of
invariants follows from the expression:
\begin{eqnarray}
\mathcal{N}\left(  \frak{g}\right)  =\dim\frak{g}-2j_{0}\left(  \frak{g}%
\right). \label{RC}
\end{eqnarray}

\section{Naturally graded nilpotent algebras of maximal index}

To any Lie algebra $\frak{r}$ we can naturally associate various
recursive series of ideals:%
\begin{eqnarray}
D^{0}\frak{r}  &  =\frak{r\supset} D^{1}\frak{r}=\left[  \frak{r}%
,\frak{r}\right]  \supset...\supset D^{k}\frak{r}=\left[  D^{k-1}%
\frak{r},D^{k-1}\frak{r}\right]  \supset...\\
C^{0}\frak{r}  &  =\frak{r\supset} C^{1}\frak{r}=\left[  \frak{r}%
,\frak{r}\right]  \supset...\supset C^{k}\frak{r}=\left[  \frak{r}%
,D^{k-1}\frak{r}\right]  \supset...
\end{eqnarray}
called respectively the derived and central descending sequence. Solvability
is given whenever the derived series is finite, i.e., if there exists a $k$
such that $D^{k}\frak{r}=0$, and nilpotency whenever the central descending
sequence is finite, i.e., if $C^{k}\frak{r}=0$ for some $k$. The dimensions of
the subalgebras in both series provide numerical invariants of the Lie
algebra. We use the notation $DS$ and $CDS$ for the dimension sequences of the
descending and central descending sequences, respectively.

Starting from the central descending sequence, we can associate a graded Lie
algebra to $\frak{r}$, which is usually denoted by $\frak{gr}\left(
\frak{r}\right)  $,
\begin{equation}
\frak{g}_{i+1}:=\frac{C^{i}\frak{r}}{C^{i+1}\frak{r}},\;i\geq0.
\end{equation}
It satisfies the condition
\begin{equation}
\left[  \frak{g}_{i},\frak{g}_{j}\right]  \subset\frak{g}_{i+j},\;1\leq i,j.
\end{equation}
A Lie algebra is said naturally graded if $\frak{r}$ and $\frak{gr}\left(
\frak{r}\right)  $ are isomorphic Lie algebras.

\begin{example}
Consider the six dimensional nilpotent Lie algebra given by the
brackets
\begin{eqnarray}
\left[  X_{2},X_{6}\right]   &  =\left[  X_{3},X_{4}\right]  =X_{1},\nonumber\\
\left[  X_{3},X_{5}\right]   &  =\left[  X_{4},X_{6}\right]  =X_{2},\nonumber\\
\left[  X_{4},X_{5}\right]   &  =X_{3},\;\left[  X_{5},X_{6}\right]  =X_{4}.
\end{eqnarray}
over the basis $\left\{  X_{1},..,X_{6}\right\}  $. It is isomorphic to the Lie algebra $A_{6,22}$ of the classification in \cite{Sh}, and satisfies
$DS=\left[  6,4,1,0\right]  $ and $CDS=\left[  6,4,3,2,1,0\right]  $. The
associated graded Lie algebra $\frak{gr}\left(  A_{6,22}\right)  $ has the
brackets
\begin{eqnarray}
\left[  X_{2},X_{6}\right]   &  =\left[  X_{3},X_{4}\right]  =X_{1},\nonumber\\
\left[  X_{3},X_{5}\right]   &  =X_{2},\nonumber\\
\left[  X_{4},X_{5}\right]   &  =X_{3},\;\left[  X_{5},X_{6}\right]  =X_{4}.
\end{eqnarray}
Therefore $A_{6,22}$ is not naturally graded, since it is not isomorphic to
$\frak{gr}\left(  A_{6,22}\right)  $. However, it can be shown that
$\frak{gr}\left(  A_{6,22}\right)  $ is a In\"{o}n\"{u}-Wigner contraction of
$A_{6,22}$.
\end{example}

Indeed, it can be shown that the graded algebras $\frak{gr}\left(
\frak{n}\right)  $ are always a contraction of the Lie algebra $\frak{n}$.
 Therefore the naturally graded nilpotent Lie algebras play a
central role in the classification of nilpotent Lie algebras.

In \cite{Wi} the solvable Lie algebras having the $n$-dimensional nilpotent Lie
algebras $\frak{n}_{n,1}$ defined by
\begin{equation}
\begin{array}
[c]{ll}%
\left[  X_{1},X_{i}\right]  =X_{i+1}, & \;2\leq i\leq n-1
\end{array}
\end{equation}
as nilradical were analyzed. The $\frak{n}_{n,1}$ algebra has maximal nilindex
and also has an abelian subalgebra of maximal dimension, generated by $\left\{
X_{2},..,X_{n}\right\}$. It can be easily verified that this
algebra is naturally graded. The question whether there exist other naturally
graded nilpotent Lie algebras of dimension $n$ and nilindex $n-1$ was answered
in \cite{Vr}.

\begin{proposition}
Let $\frak{n}$ be a naturally graded nilpotent Lie algebra of dimension $n$
and central descending sequence $CDS=\left[  n,n-2,n-3,..,2,1,0\right]  $.
Then $\frak{n}$ is isomorphic to $\frak{n}_{n,1}$ if $n$ is odd, and
isomorphic  to $\frak{n}_{n,1}$ or $Q_{2n}$ if $n$ is even, where

\begin{enumerate}
\item $\frak{n}_{n,1}:$
\[%
\begin{array}
[c]{ll}%
\left[  X_{1},X_{i}\right]  =X_{i+1}, & \;2\leq i\leq n-1
\end{array}
\]
over the basis $\left\{  X_{1},..,X_{n}\right\}  ,$

\item $Q_{2m} (m\geq 3):$
\[%
\begin{array}
[c]{ll}%
\left[  X_{1},X_{i}\right]  =X_{i+1}, & \;2\leq i\leq2m-2\\
\left[  X_{k},X_{2n+1-k}\right]  =\left(  -1\right)  ^{k}X_{2m}, & 2\leq k\leq
m.
\end{array}
\]
over the basis $\left\{  X_{1},..,X_{2m}\right\}  $.
\end{enumerate}
\end{proposition}

We mention that for $n=2m=4$ the Lie algebras $\frak{n}_{4,1}$ and
$Q_{4}$ are isomorphic. In higher dimensions, they are no more isomorphic, but related by a generalized In\"on\"u-Wigner contraction. More precisely, consider the automorphism of $Q_{2n}$ given by the matrix
\begin{equation*}
\left(  X_{1}^{\prime},..,X_{2n}^{\prime}\right)^{T}  =\left(
\begin{array}
[c]{cccccc}%
1 & 0 & 0 & ... & 0 & 0\\
1 & \varepsilon & 0 & ... & 0 & 0\\
0 & 0 & \varepsilon & ... & 0 & 0\\
\vdots & \vdots & \vdots & \ddots & \vdots & \vdots\\
0 & 0 & 0 & 0 & \varepsilon & 0\\
0 & 0 & 0 & 0 & 0 & \varepsilon
\end{array}
\right)  \left(
\begin{array}
[c]{c}%
X_{1}\\
X_{2}\\
\vdots\\
\vdots\\
X_{2n-1}\\
X_{2n}%
\end{array}
\right)  .
\end{equation*}
Over the new basis the brackets of $Q_{2n}$ are expressed as:%
\begin{equation}
\begin{array}
[c]{ll}%
\left[  X_{1}^{\prime},X_{i}^{\prime}\right]  =X_{i+1}^{\prime}, & \;2\leq
i\leq2n-1\\
\left[  X_{k}^{\prime},X_{2n+1-k}^{\prime}\right]  =\left(  -1\right)
^{k}\varepsilon X_{2n}^{\prime}, & 2\leq k\leq n.
\end{array}
\end{equation}
In the limit $\varepsilon\rightarrow0$ we obtain the Lie algebra
$\frak{n}_{2n,1}$. Inspite of this fact, the nilpotent Lie algebras $Q_{2n}$ and $\frak{n}_{n,1}$
have a very different behaviour. While the second has a number of Casimir
operators which depends on the dimension \cite{Wi}, the algebra $Q_{2n}$ has a
fixed number of invariants for any $n$.

\begin{proposition}
For any $n\geq3$ the nilpotent Lie algebra $Q_{2n}$ has exactly $2$ Casimir
operators, given by the symmetrization of the following functions:
\begin{eqnarray}
I_{1} &  =x_{2n}\\
I_{2} &  =x_{1}x_{2n}+x_{3}x_{2n-1}+\sum_{k=4}^{n}\left(  -1\right)
^{k+1}x_{k}x_{2n+2-k}+\frac{\left(  -1\right)  ^{n}}{2}x_{n+1}^{2}.\label{qua}
\end{eqnarray}
\end{proposition}

\begin{proof}
The Maurer-Cartan equations of the algebra $Q_{2n}$ are
\begin{eqnarray}
d\omega_{1} &  =d\omega_{2}=0,\nonumber\\
d\omega_{j+1} &  =\omega_{1}\wedge\omega_{j},\;2\leq j\leq2n-2\nonumber\\
d\omega_{2n} &  =\sum_{k=2}^{n}\left(  -1\right)  ^{k}\omega_{k}\wedge
\omega_{2n+1-k}.
\end{eqnarray}
The 2-form $\omega=d\omega_{2n}$ is of maximal rank, therefore
$j_{0}\left(  Q_{2n}\right)  =n-1$ and by formula (\ref{RC}) we have $\mathcal{N}\left(
Q_{2n}\right)  =2$. Clearly the generator of the centre is one Casimir
operator of the algebra. In order to determine the other independent
invariant, we have to solve the system (\ref{sys}):
\begin{eqnarray}
\widehat{X}_{1}F &  :=\sum_{j=2}^{2n-2}x_{j+1}\frac{\partial F}{\partial
x_{j}}=0 \label{eq1}\\
\widehat{X}_{j}F &  :=-x_{j+1}\frac{\partial F}{\partial x_{1}}+\left(
-1\right)  ^{j}x_{2n}\frac{\partial F}{\partial x_{2n+1-j}}=0 \label{eq2}\\
\widehat{X}_{2n-1}F &  :=-x_{2n}\frac{\partial F}{\partial x_{2}}=0,\label{eq3}
\end{eqnarray}
where $2\leq j\leq2n-2$. Equation (\ref{eq3}) implies that $\frac{\partial F}{\partial
x_{2}}=0$. For any fixed $2\leq j\leq 2n-2$, the function $x_{1}x_{2n}+(-1)^{j}x_{j+1}x_{2n+1-j}$ is a solution of equation (\ref{eq2}). If we consider the function  $C=x_{1}x_{2n}+x_{3}x_{2n-1}+\sum_{k=4}^{n}\left(  -1\right)
^{k+1}x_{k}x_{2n+2-k}+\frac{\left(  -1\right)  ^{n}}{2}x_{n+1}^{2}$, for any $j\geq 3$ the following identity is satisfied:
\begin{equation}
x_{j+2}\frac{\partial C}{\partial x_{j+1}}+x_{2n+1}\frac{\partial C}{\partial x_{2n-j}}=0.
\end{equation}
This implies that $\widehat{X}_{1}(C)=0$, and therefore that $C$ is an invariant of the algebra. \newline The Casimir operator follows at once replacing $x_{i}$ by $X_{i}$. Observe in particular that $C$ coincides with its symmetrization.
\end{proof}

\bigskip

\subsection{Solvable Lie algebras with fixed nilradical}

Any solvable Lie algebra $\frak{r}$ over the real or complex field admits a
decomposition%
\begin{equation}
\frak{r}=\frak{t}\overrightarrow{\oplus}\frak{n}
\end{equation}
satisfying the relations%
\begin{equation}
\left[  \frak{t,n}\right]  \subset\frak{n,\;}\left[  \frak{n},\frak{n}\right]
\subset\frak{n},\;\left[  \frak{t},\frak{t}\right]  \subset\frak{n,}%
\end{equation}
where $\frak{n}$ is the maximal nilpotent ideal of $\frak{r}$, called the
nilradical, and $\overrightarrow{\oplus}$ denotes the semidirect product. It
was proven in \cite{Mu63} that the dimension of the nilradical satisfies
the following inequality%
\begin{equation}
\dim\frak{n}\geq\frac{1}{2}\dim\frak{r}.
\end{equation}

Applying the Jacobi identity to any elements $X\in\frak{t}$, $Y_{1},Y_{2}%
\in\frak{n}$, we obtain that%
\begin{equation}
\left[  X,\left[  Y_{1},Y_{2}\right]  \right]  +\left[
Y_{2},\left[ X,Y_{1}\right]  \right]  +\left[  Y_{1},\left[
Y_{2},X\right]  \right] =0,
\end{equation}
i.e., $ad\left(  X\right)  $ acts as a derivation of the nilpotent
algebra $\frak{n}$. Since the elements $X\notin\frak{n}$, these
derivations are not nilpotent, and given a basis $\left\{
X_{1},..,X_{n}\right\}  $ of $\frak{t}$ and arbitrary scalars
$\alpha_{1},..,\alpha_{n}\in\mathbb{R-}\left\{ 0\right\}  $, we
have that
\begin{equation}
\left(  \alpha_{1}ad\left(  X_{1}\right)  +..+\alpha_{n}ad\left(
X_{n}\right)  \right)  ^{k}\neq0,\quad k\geq1,\label{NI}%
\end{equation}
that is, the matrix $\alpha_{1}ad\left(  X_{1}\right)
+..+\alpha_{n}ad\left( X_{n}\right)  $ is not nilpotent. We say
that the elements $X_{1},..,X_{n}$ are nil-independent
\cite{Mu63}. This fact imposes a first restriction on the
dimension of a solvable Lie algebra having a given nilradical,
namely, that $\dim\frak{r}$ is bounded by the maximal number of
nil-independent derivations of the nilradical.. Therefore, the
classification of solvable Lie algebras reduces to the problem of
finding all non-equivalent extensions determined by a set of
nil-independent derivations. The equivalence of extensions is
considered under the transformations of the type
\begin{equation}
X_{i}\mapsto a_{ij}X_{j}+b_{ik}Y_{k},\quad Y_{k}\mapsto R_{kl}Y_{l}, \label{TA}
\end{equation}
where $(a_{ij})$ is an invertible $n\times n$ matrix, $(b_{ik})$ is a $n\times \dim\frak{n}$ matrix and $(R_{kl})$ is an automorphism of the nilradical $\frak{n}$.

\section{Derivations of $Q_{2n}$}
In this section we determine the algebra of derivations of $Q_{2n}$, in order to obtain all equivalence classes
of extensions by non-nilpotent derivations.

\begin{proposition}
Any outer derivation $f\in Der\left(  Q_{2n}\right)  $ has the form
\begin{eqnarray}
\fl f\left(  X_{1}\right)   &  =\lambda_{1}X_{1}+f_{1}^{2n}X_{2n}\nonumber\\
\fl f\left(  X_{2}\right)   &  =\lambda_{2}X_{2}+\sum_{k=2}^{n-1}f_{2}%
^{2k+1}X_{2k+1}\nonumber\\
\fl f\left(  X_{2+t}\right)   &  =\left(  t\lambda_{1}+\lambda_{2}\right)
X_{2+t}+\sum_{k=2}^{\left[  \frac{2n-3-t}{2}\right]  }f_{2}^{2k+1}%
X_{2k+1+t},\;1\leq t\leq2n-4\nonumber\\
\fl f\left(  X_{2n-1}\right)   &  =\left(  \left(  2n-3\right)  \lambda
_{1}+\lambda_{2}\right)  X_{2n-1}\nonumber\\
\fl f\left(  X_{2n}\right)   &  =\left(  \left(  2n-3\right)  \lambda_{1}%
+2\lambda_{2}\right)  X_{2n}.
\end{eqnarray}
In particular $\dim Der\left(  Q_{2n}\right)/IDer(Q_{2n}  =n+1$.
\end{proposition}

\begin{proof}
For convenience we denote for any $1\leq i\leq2n$%
\begin{equation}
f\left(  X_{i}\right)  =\sum_{j=1}^{2n}f_{i}^{j}X_{j},
\end{equation}
where the $f_{i}^{j}$ are scalars. Since any derivation maps central elements
onto central elements, we have that
\begin{equation}
f\left(  X_{2n}\right)  =f_{2n}^{2n}X_{2n}.
\end{equation}
The condition%
\begin{equation}
\left[  f\left(  X_{1}\right)  ,X_{2}\right]  +\left[  X_{1},f\left(
X_{2}\right)  \right]  =f\left(  X_{3}\right) \label{D1}
\end{equation}
shows that
\begin{equation}
f\left(  X_{3}\right)  =\left(  f_{1}^{1}+f_{2}^{2}\right)  X_{3}+\sum
_{k=3}^{2n-2}f_{2}^{k}X_{k+1}-f_{1}^{2n}X_{2n}.
\end{equation}
Since $X_{2+t}=ad\left(  X_{1}\right)  ^{t}\left(  X_{2}\right)  $ for $1\leq
t\leq2n-3$, iteration of equation (\ref{D1}) shows that
\begin{equation}
f\left(  X_{2+t}\right)  =\left(  tf_{1}^{1}+f_{2}^{2}\right)  X_{2+t}%
+\sum_{k=3}^{2n-1-t}f_{2}^{k}X_{k+t}+\left(  -1\right)  ^{t}f_{1}%
^{2n-1-t}X_{2n}.
\end{equation}
The condition
\begin{equation*}
\left[  f\left(  X_{1}\right)  ,X_{2n-1}\right]  +\left[  X_{1},f\left(
X_{2n-1}\right)  \right]  =0
\end{equation*}
implies that $f_{1}^{2}=0$. In particular it follows that
\begin{equation}
f\left(  X_{2n-1}\right)  =\left(  \left(  2n-3\right)  f_{1}^{1}+f_{2}%
^{2}\right)  X_{2n}.
\end{equation}
We now evaluate the Leibniz condition on the pairs $\left(  X_{2}%
,X_{2+t}\right)  $ for $1\leq t\leq2n-4$:%
\begin{equation*}
\fl \left[  f\left(  X_{2}\right)  ,X_{2+t}\right]  +\left[  X_{2},f\left(
X_{2+t}\right)  \right]  =f_{2}^{2n-1-t}\left(  1-\left(  -1\right)
^{t}\right)  X_{2n}+f_{2}^{1}X_{3+t}=0,
\end{equation*}
from which we deduce that $f_{2}^{1}=0$ and
\begin{eqnarray}
f_{2}^{1}  & =0\nonumber\\
f_{2}^{2n-1-t}  & =0,\;t=1,3,..,2n-5.
\end{eqnarray}
It can be easily show that for all $k=2,..,n$ we obtain
\[
\left[  f\left(  X_{k}\right)  ,X_{2n+1-k}\right]  =\left(  \left(
2n-3\right)  f_{1}^{1}+2f_{2}^{2}\right)  X_{2n}=f\left(  X_{2n}\right).
\]
The remaining brackets give no new conditions on the coefficients $f_{i}^{j}$.
Therefore any derivation $f$ has the form:%
\begin{eqnarray}
\fl f\left(  X_{1}\right)    =f_{1}^{1}X_{1}+\sum_{k=3}^{2n}f_{1}^{k}X_{2n}\nonumber\\
\fl f\left(  X_{2}\right)     =f_{2}^{2}X_{2}+\sum_{k=2}^{n-1}f_{2}%
^{2k+1}X_{2k+1}+f_{2}^{2n}X_{2n}\nonumber\\
\fl f\left(  X_{2+t}\right)  =\left(  tf_{1}^{1}+f_{2}^{2}\right)
X_{2+t}+\sum_{k=2}^{\left[  \frac{2n-3-t}{2}\right]  }f_{2}^{2k+1}%
X_{2k+1+t}+\left(  -1\right)  ^{t}f_{1}^{2n-1-t}X_{2n},\;1\leq t\leq2n-4\nonumber\\
\fl f\left(  X_{2n-1}\right)  =\left(  \left(  2n-3\right)  \lambda
_{1}+\lambda_{2}\right)  X_{2n-1}\nonumber\\
\fl f\left(  X_{2n}\right)    =\left(  \left(  2n-3\right)  \lambda_{1}%
+2\lambda_{2}\right)  X_{2n}.
\end{eqnarray}
Since there are $3n$ parameters, we conclude that
\[
\dim Der\left(  Q_{2n}\right)  =3n.
\]
For any of these parameters we define the following derivations:
\begin{eqnarray}
\fl F_{1}^{1}\left(  X_{1}\right)     =X_{1},\;F_{1}^{1}\left(  X_{j}\right)
=\left(  j-2\right)  X_{j},\;F_{1}^{1}\left(  X_{2n}\right)  =\left(
2n-3\right)  X_{2n},\;3\leq j\leq2n-1\nonumber\\
\fl F_{2}^{2}\left(  X_{2}\right)     =X_{2},\;F_{2}^{2}\left(  X_{j}\right)
=X_{j},\;F_{2}^{2}\left(  X_{2n}\right)  =2X_{2n},\;3\leq j\leq2n-1\nonumber\\
\fl F_{1}^{k}\left(  X_{1}\right)     =X_{k},\;F_{1}^{k}\left(  X_{2n+2-k}%
\right)  =\left(  -1\right)  ^{k}X_{2n},\;3\leq j\leq2n-1\nonumber\\
\fl F_{1}^{2n}\left(  X_{1}\right)    =X_{2n},\nonumber\\
\fl F_{1}^{2k+1}\left(  X_{2}\right)  =X_{2k+1},\;F_{2}^{2k+1}\left(
X_{2+t}\right)  =X_{2k+1+t},\;1\leq t\leq2\left(  n-1-k\right)  ,\;2\leq k\leq
n-1\nonumber\\
\fl F_{1}^{2n}\left(  X_{2}\right)    =X_{2n},\;
\end{eqnarray}
To obtain the outer derivations, we have to determine all adjoint operators
$ad\left(  X\right)  $ for $X\in Q_{2n}$: It can be easily seen that following
relations hold:%
\[
adX_{1}=F_{2}^{3},\;adX_{2}=F_{1}^{3},\;adX_{k}=F_{1}^{k}\;\left(  3\leq
k\leq2n-2\right)  ,\;adX_{2n-1}=F_{2}^{2n}.
\]
Therefore there are $n+1$ outer derivations, corresponding to the derivations
$\left\{  F_{1}^{1},F_{1}^{2n},F_{2}^{2},F_{2}^{2k+1}\right\}  _{1\leq
k\leq2n-1}$.
\end{proof}

\begin{corollary}
Any non-nilpotent outer derivation $F$ of $Q_{2n}$ is of the form
\begin{equation}
F=\alpha_{1}F_{1}^{1}+\alpha_{2}F_{2}^{2}+\sum_{k} \beta_{k}F_{2}^{2k+1}+\beta_{n}%
F_{1}^{2n} \label{De1}
\end{equation}
where either $\alpha_{1}\neq0$ or $\alpha_{2}\neq0$.
\end{corollary}

\section{Solvable Lie algebras with $Q_{2n}$-nilradical}

In this section we apply the preceding results on derivations and equation (\ref{De1}) to classify the solvable real and complex Lie algebras having a  nilradical isomorphic to the graded algebra $Q_{2n}$.

\begin{proposition}
Any solvable Lie algebra $\frak{r}$ with nilradical isomorphic to $Q_{2n}$ has
dimension $2n+1$ or $2n+2$.
\end{proposition}

The proof is an immediate consequence of corollary 1.

\begin{proposition}
Any solvable Lie algebra of dimension $2n+1$ with nilradical isomorphic
to $Q_{2n}$ is isomorphic to one of the following algebras:

\begin{enumerate}
\item $\frak{r}_{2n+1}\left(  \lambda_{2}\right)  :$%
\[%
\begin{array}
[c]{llll}%
\left[  X_{1},X_{k}\right]  =X_{k+1}, & 2\leq k\leq2n-2 &  & \\
\left[  X_{k},X_{2n+1-k}\right]  =\left(  -1\right)  ^{k}X_{2n}, & 2\leq k\leq n &
& \\
\left[  Y,X_{1}\right]  =X_{1}, &  &  & \\
\left[  Y,X_{k}\right]  =\left(  k-2+\lambda_{2}\right)  X_{k}, & 2\leq
k\leq2n-1 &  & \\
\left[  Y,X_{2n}\right]  =\left(  2n-3+2\lambda_{2}\right)  X_{2n}. &  &  &
\end{array}
\]

\item $\frak{r}_{2n+1}\left(  2-n,\varepsilon\right)  $%
\[%
\begin{array}
[c]{llll}%
\left[  X_{1},X_{k}\right]  =X_{k+1}, & 2\leq k\leq2n-2 &  & \\
\left[  X_{k},X_{2n+1-k}\right]  =\left(  -1\right)  ^{k}X_{2n}, & 2\leq k\leq n &
& \\
\left[  Y,X_{1}\right]  =X_{1}+\varepsilon X_{2n}, & \varepsilon=-1,0,1 &  &
\\
\left[  Y,X_{k}\right]  =\left(  k-n\right)  X_{k}, & 2\leq k\leq2n-1 &  & \\
\left[  Y,X_{2n}\right]  =X_{2n}. &  &  &
\end{array}
\]

\item $\frak{r}_{2n+1}\left(  \lambda_{2}^{5},..,\lambda_{2}^{2n-1}\right)  $%
\[%
\begin{array}
[c]{llll}%
\left[  X_{1},X_{k}\right]  =X_{k+1}, & 2\leq k\leq2n-2 &  & \\
\left[  X_{k},X_{2n+1-k}\right]  =\left(  -1\right)  ^{k}X_{2n}, & 2\leq k\leq n &
& \\
\left[  Y,X_{2+t}\right]  =X_{2+t}+\sum_{k=2}^{\left[  \frac{2n-3-t}%
{2}\right]  }\lambda_{2}^{2k+1}X_{2k+1+t}, & 0\leq t\leq2n-6 &  & \\
\left[  Y,X_{2n-k}\right]  =X_{2n-k}, & k=1,2,3 &  & \\
\left[  Y,X_{2n}\right]  =2X_{2n}. &  &  &
\end{array}
\]
Moreover, the first nonvanishing parameter $\lambda_{2}^{2k+1}$ can be
normalized to $1$.
\end{enumerate}
\end{proposition}

\begin{proof}
Let $F=\alpha_{1}F_{1}^{1}+\alpha_{2}F_{2}^{2}+\sum_{k}\beta_{k}F_{2}^{2k+1}+\beta_{n}%
F_{1}^{2n}$ be a non-nilpotent derivation.

\begin{enumerate}
\item  Let $\alpha_{1}\neq0$. A scaling change allows us to suppose that
$\alpha_{1}=1$. By a change of the type%
\begin{eqnarray}
X_{2+t}^{\prime}  & =X_{2+t}+\sum_{k=2}^{\left[  \frac{2n-3-t}{2}\right]
}\gamma_{k}X_{2k+1+t},\;0\leq t\leq2n-4\nonumber\\
X_{i}^{\prime}  & =X_{i},\;i=1,2n-1,2n,
\end{eqnarray}
we put to zero first $f_{2}^{2n-1},$ then $f_{2}^{2n-3}$ etc. up to $f_{2}%
^{5}$. This shows that the extension given by $F$ is equivalent to the
extension defined by
\begin{equation}
F^{\prime}=F_{1}^{1}+\alpha_{2}F_{2}^{2}+\beta_{n}F_{1}^{2n}.
\end{equation}
If further $\alpha_{2}\neq2-n$, then the change of basis
\begin{equation}
X_{1}^{\prime}=X_{1}+\frac{f_{1}^{2n}}{2\left(  n-2+\alpha_{2}\right)  }X_{2n}%
\end{equation}
allows us to suppose $f_{1}^{2n}=0$. Therefore the derivation is diagonal and
has eigenvalues
\begin{equation}
\Delta=\left(  1,\alpha_{2},1+\alpha_{2},..,2n-3+\alpha_{2},2n-3+2\alpha
_{2}\right)
\end{equation}
over the ordered basis $\left\{  X_{1},..,X_{2n}\right\}  $ of $Q_{2n}$. We
obtain the solvable Lie algebras $\frak{r}_{2n+1}\left(  \alpha_{2}\right)  $.
However, if $\alpha_{2}=2-n$ and $f_{1}^{2n}\neq0$, then it cannot be removed.
The only possibility is to consider scaling transformations. Over $\mathbb{F}=\mathbb{R}$ this allows us to put $f_{1}^{2n}$ equal
to $1$ if $f_{1}^{2n}>0$ or $f_{1}^{2n}=-1$ if $f_{1}^{2}<0$, while over $\mathbb{F}=\mathbb{C}$ we can always normalize $f_{1}^{2n}$ to $1$. This gives the Lie algebras $\frak{r}_{2n+1}\left(  2-n,\varepsilon\right)$\footnote{The Lie algebras $\frak{r}_{2n+1}\left(  2-n,-1\right)$ and $\frak{r}_{2n+1}\left(  2-n,1\right)$ being isomorphic over $\mathbb{C}$.}. In addition,
if $\alpha_{2}=2-n$ but  $f_{1}^{2n}=0$, we obtain the Lie algebra
$\frak{r}_{2n+1}\left(  2-n\right)  $.

\item  Let us suppose now that $\alpha_{1}=0$. By nil-independence we have
$\alpha_{2}\neq0$ and by scaling transformation we can suppose that $\alpha_{2}=1$. The change of
basis
\begin{equation}
X_{1}^{\prime}=X_{1}-\frac{1}{2}f_{1}^{2n}X_{2n}%
\end{equation}
allows us to put $f_{1}^{2n}$ to zero. Now the parameters $f_{2}^{2k+1}$
$\left(  2\leq k\leq n-1\right)  $ cannot be removed, so that unless all
vanish, the derivation $F$ is not diagonal. However, the first non-vanishing
parameter $f_{2}^{2k+1}$ can always be normalized to $1$ by a scaling
change of basis. This case provides the family of algebras $\frak{r}_{2n+1}\left(
f_{2}^{5},..,f_{2}^{2n-1}\right)  .$
\end{enumerate}
\end{proof}

Finally, if we add the two nil-independent elements, there is only one possibility:

\begin{proposition}
For any $n\geq3$ there is only one solvable Lie algebra $\frak{r}_{2n+2}$ of
dimension $2n+2$ having a nilradical isomorphic to $Q_{2n}:$%
\[%
\begin{array}
[c]{llll}%
\left[  X_{1},X_{k}\right]  =X_{k+1}, & 2\leq k\leq2n-2 &  & \\
\left[  X_{k},X_{2n+1-k}\right]  =\left(  -1\right)  ^{k}X_{2n}, & 2\leq k\leq n &
& \\
\left[  Y_{1},X_{k}\right]  =kX_{k} & 1\leq k\leq2n-1 &  & \\
\left[  Y_{1},X_{2n}\right]  =\left(  2n+1\right)  X_{2n}, &  &  & \\
\left[  Y_{2},X_{k}\right]  =X_{k}, & 2\leq k\leq2n-1 &  & \\
\left[  Y_{2},X_{2n}\right]  =2X_{2n}. &  &  &
\end{array}
\]
\end{proposition}

\section{The generalized Casimir invariants}

We now consider the solvable Lie algebras obtained in the previous section, and compute their
generalized Casimir invariants.

\begin{theorem}
The Lie algebras $\frak{r}\left(  \lambda_{2}\right)  ,\frak{r}\left(
2-n,\varepsilon\right)  $ and $\frak{r}\left(   \lambda_{2}%
^{5},..,\lambda_{2}^{2n-1}\right)  $ have one invariant for any dimension.
They can be chosen as follows:

\begin{enumerate}
\item $\frak{r}_{2n+1}\left(  \lambda_{2}\right)  $%
\begin{equation}
J=I_{2}x_{2n}^{-\alpha},\;\alpha=\frac{2n-2+2\lambda_{2}}{2n-3+2\lambda_{2}},
\end{equation}

\item $\frak{r}_{2n+1}\left(  2-n,\varepsilon\right)  $%
\begin{equation}
J=\frac{1}{x_{2n}^{2}}I_{2}-\varepsilon\ln\left(  x_{2n}\right)  ,
\end{equation}

\item $\frak{r}_{2n+1}\left(  \lambda_{2}^{5},..,\lambda_{2}^{2n-1}\right)  $%
\begin{equation}
J=\frac{I_{2}}{x_{2n}},
\end{equation}
where in all cases
\begin{equation}
I_{2}=x_{1}x_{2n}+x_{3}x_{2n-1}+\sum_{j=4}^{n}\left(  -1\right)  ^{j}%
x_{j}x_{2n+2-j}+\frac{\left(  -1\right)  ^{n+1}}{2}x_{n+1}^{2}.
\end{equation}
\end{enumerate}
\end{theorem}

\begin{proof}
Using the Maurer-Cartan equations of the solvable Lie algebras above, it is straightforward to verify that in all cases we have $\mathcal{N}(\frak{r})=1$. If moreover the derivation $F$ defining the extension of $Q_{2n}$ acts nontrivially on the centre $X_{2n}$, then clearly the invariants are independent on the variable $y$ associated to the generator $Y$. To find the invariants of the solvable algebras reduces then to solve the equation $\widehat{Y}F=0$, taking into account the invariants $I_{1}=x_{2n}$ and
\begin{equation}
I_{2}=x_{1}x_{2n}+x_{3}x_{2n-1}+\sum_{j=4}^{n}\left(  -1\right)  ^{j}%
x_{j}x_{2n+2-j}+\frac{\left(  -1\right)  ^{n+1}}{2}x_{n+1}^{2}.
\end{equation}
obtained in proposition 2.

\begin{enumerate}

\item $\frak{r}\left(  \lambda_{2}\right)$. \newline The equation to be solved is
\begin{equation}
\fl \widehat{Y}F:=x_{1}\frac{\partial F}{\partial x_{1}}+\sum_{k=2}^{2n-1}\left(
k-2+\lambda_{2}\right)  x_{k}\frac{\partial F}{\partial x_{k}}+\left(
2n-3+2\lambda_{2}\right)  x_{2n}\frac{\partial F}{\partial x_{2n}}=0. \label{ev}
\end{equation}
It can be easily verified that
\begin{eqnarray*}
\widehat{Y}\left(  I_{1}\right)    & =\left(  2n-3+2\lambda_{2}\right)
I_{1}\\
\widehat{Y}\left(  I_{2}\right)    & =\left(  2n-2+2\lambda_{2}\right)
I_{2}.\nonumber
\end{eqnarray*}
This means that the Casimir operators of the nilradical are semi-invariants of the solvable extension. This fact always holds for diagonal derivations \cite{Wi1,Ca3,Wi,C48}. Observe that if $2n-3+2\lambda_{2}=0$, then $J=x_{2n}$ is already the
invariant of the algebra, while for $2n-2+2\lambda_{2}=0$ the function $I_{2}$
is a Casimir operator of $\frak{r}_{2n+1}\left(  2-n\right)  $. If neither of
$I_{1}$ or $I_{2}$ is a solution of (\ref{ev}), then, considering
$I_{1}$ and $I_{2}$ as new variables $u$ and $v$, we take the differential
equation
\begin{equation}
\frac{\partial\Phi}{\partial u}+\frac{\left(  2n-2+2\lambda_{2}\right)
v}{\left(  2n-3+2\lambda_{2}\right)  u}\frac{\partial\Phi}{\partial v}=0, \label{dif}
\end{equation}
with general solution
\begin{equation}
\Phi\left(  \frac{u^{2n-2+2\lambda_{2}}}{v^{2n-3+2\lambda_{2}}}\right)  .
\end{equation}
Therefore the invariant of $\frak{r}_{2n+1}\left(  \lambda_{2}\right)  $ can
be taken as
\begin{equation}
J=I_{2}x_{2n}^{-\alpha},\;\alpha=\left(  \frac{2n-2+2\lambda_{2}%
}{2n-3+2\lambda_{2}}\right).
\end{equation}

\item $\frak{r}_{2n+1}\left(2-n,\epsilon\right)$.\newline
In this case we have
\begin{equation}
\widehat{Y}F:=(x_{1}+\varepsilon x_{2n})\frac{\partial F}{\partial x_{1}}%
+\sum_{k=2}^{2n-1}\left(  k-n\right)  x_{k}\frac{\partial F}{\partial x_{k}%
}+x_{2n}\frac{\partial F}{\partial x_{2n}}=0. \label{ev2}
\end{equation}
Since the action is not diagonal when $\varepsilon\neq0$, $\widehat{Y}\left(
I_{2}\right)  $ will not be a multiple of $I_{2}$. However, replacing $I_{2}$ by $I_{2}x_{2n}^{-2}$, we obtain
\begin{equation*}
\widehat{Y}\left(  I_{2}x_{2n}^{-2}\right)  =\varepsilon.
\end{equation*}
Since $\widehat{Y}(\ln(x_{2n}))=1$, adding the logarithm $-\varepsilon\ln\left(
x_{2n}\right)$, the function
\begin{equation}
I_{2}x_{2n}^{-2}-\varepsilon\ln\left(  x_{2n}\right)
\end{equation}
is a solution of (\ref{ev2}), and can be taken as invariant of the algebra.

\item $\frak{r}_{2n+1}\left(  \lambda_{2}^{5},..,\lambda_{2}^{2n-1}\right)  $.\newline
For the families the equation to be solved is
\begin{equation}
\widehat{Y}F:=\sum_{k=2}^{2n-1}x_{k}\frac{\partial F}{\partial x_{k}}%
+2x_{2n}\frac{\partial F}{\partial x_{2n}}=0.
\end{equation}
After some calculation it follows that
\begin{eqnarray}
\widehat{Y}\left(  I_{1}\right)    & =2I_{1}\nonumber\\
\widehat{Y}\left(  I_{2}\right)    & =2I_{2},\nonumber
\end{eqnarray}
so that applying the same method as in (\ref{dif}), the invariant can be chosen as
\begin{equation}
J=\frac{I_{2}}{I_{1}}.
\end{equation}
\end{enumerate}

\end{proof}

As expected, most of the solvable algebras have harmonics as invariants (see \cite{Wi} for the invariants in the $\frak{n}_{n,1}$ case). Only for special values of the parameters classical Casimir operators  are obtained. It should be noted that no rational basis of invariants of $\frak{r}_{2n+1}(2-n,\epsilon)$ can be obtained for $\epsilon\neq 0$.

\begin{proposition}
The Lie algebra $\frak{r}_{2n+2}$ has no invariants for any $n\geq3$.
\end{proposition}

If $\left\{  \omega_{1},..,\omega_{2n},\theta_{1},\theta_{2}\right\}  $
denotes a dual basis to $\left\{  X_{1},..,X_{2n},Y_{1},Y_{2}\right\}  $, then
the Maurer-Cartan equations have the form%
\begin{eqnarray}
d\omega_{1}  =\omega_{1}\wedge\theta_{1}\nonumber\\
d\omega_{2}   =2\omega_{2}\wedge\theta_{1}+\omega_{2}\wedge\theta_{2}\nonumber\\
d\omega_{k}   =\omega_{1}\wedge\omega_{k-1}+k\omega_{k}\wedge\theta
_{1}+\omega_{k}\wedge\theta_{2},\;3\leq k\leq2n-1\nonumber\\
d\omega_{2n}  =\sum_{k=2}^{n}\left(  -1\right)  ^{k}\omega_{k}\wedge
\omega_{2n+1-k}+\left(  2n+1\right)  \omega_{2n}\wedge\theta_{1}+2\omega
_{2n}\wedge\theta_{2}\nonumber\\
d\theta_{1}   =d\theta_{2}=0.
\end{eqnarray}
Taking the form $\xi=d\omega_{1}+d\omega_{2n}$ and computing the $n^{th}$
wedge product we obtain
\begin{equation}
\bigwedge^{n}\xi=\pm\left(  2n\right)  n!\omega_{1}\wedge...\wedge\omega
_{2n}\wedge\theta_{1}\wedge\theta_{2}\neq0,
\end{equation}
and by formula (\ref{RC}) the Lie algebra has no invariants.

\section{Geometric properties of solvable Lie algebras with $Q_{2n}$-radical}

In view of the last proposition, which shows that the Lie algebra $\frak{r}_{2n+2}$ is endowed with an exact symplectic structure, it is natural to ask whether the other solvable Lie algebras with nilradical isomorphic to $Q_{2n}$ and dimension $2n+1$ also have special geometrical properties. Specifically, we analyze the existence of linear contacts forms on these algebras, i.e., 1-forms $\omega\in\frak{r}_{2n+1}^{*}$ such that $\omega\wedge(\bigwedge^{n} d\omega)\neq 0$. This type of geometrical structure has been shown to be of interest for the analysis of differential equations and also for dynamical systems \cite{Kru,Ree}.

\begin{proposition}
Let $n\geq3$. Any solvable Lie algebra $\frak{r}$ with nilradical isomorphic to $Q_{2n}$, with the exception of $\frak{r}_{2n+1}(2-n,0)$, is endowed with a linear contact form.
\end{proposition}

\begin{proof}
Let $\left\{\omega_{1},..,\omega_{2n},\theta\right\}$ be a dual basis of $\left\{X_{1},..,X_{2n},Y\right\}$.

\begin{enumerate}
\item  The Maurer-Cartan equations of $\frak{r}_{2n+1}\left(  \lambda
_{2}\right)  $ are
\begin{eqnarray}
d\omega_{1}   =\omega_{1}\wedge\theta\nonumber\\
d\omega_{2}   =\lambda_{2}\omega_{2}\wedge\theta\nonumber\\
d\omega_{k}   =\omega_{1}\wedge\omega_{k-1}+\left(  k-2+\lambda_{2}\right)
\omega_{k}\wedge\theta,\;3\leq k\leq2n-1\nonumber\\
d\omega_{2n}  =\sum_{k=2}^{n}\left(  -1\right)  ^{k}\omega_{k}\wedge
\omega_{2n+1-k}+\left(  2n-3+2\lambda_{2}\right)  \omega_{2n}\wedge\theta\nonumber\\
d\theta  =0.
\end{eqnarray}
Taking $\omega=\omega_{1}+\omega_{2n}$, the exterior product gives
\begin{equation}
\omega\wedge\left(  \bigwedge^{n}d\omega\right)  =2n\left(  n-1\right)
!\left(  \lambda_{2}+n-2\right)  \omega_{1}\wedge...\wedge\omega_{2n}%
\wedge\theta\neq0.
\end{equation}

\item  The Maurer-Cartan equations of $\frak{r}_{2n+1}\left(  2-n,\varepsilon
\right)  $ are
\begin{eqnarray}
d\omega_{1}   =\omega_{1}\wedge\theta\nonumber\\
d\omega_{2}   =\omega_{2}\wedge\theta\nonumber\\
d\omega_{k}   =\omega_{1}\wedge\omega_{k-1}+\left(  k-n\right)  \omega
_{k}\wedge\theta,\;3\leq k\leq2n-1\nonumber\\
d\omega_{2n}  =\sum_{k=2}^{n}\left(  -1\right)  ^{k}\omega_{k}\wedge
\omega_{2n+1-k}+\omega_{2n}\wedge\theta+\varepsilon\omega_{1}\wedge\theta\nonumber\\
d\theta  =0.
\end{eqnarray}
Taking again the 1-form  $\omega=\omega_{1}+\omega_{2n}$, we obtain that
\begin{equation}
\omega\wedge\left(  \bigwedge^{n}d\omega\right)  =\varepsilon n!\omega
_{1}\wedge...\wedge\omega_{2n}\wedge\theta.
\end{equation}
Thus $\omega$ is a contact form whenever $\varepsilon\neq0$. It is not
difficult to show that for $\varepsilon=0$ there is no linear contact form.

\item  For $\frak{r}_{2n+1}\left(  \lambda_{2}^{5},..,\lambda_{2}%
^{2n-1}\right)  $, the Maurer-Cartan equations are quite complicated, due to
the number of parameters $\lambda_{2}^{k}$. However, in order to find a contact form we can restrict ourselves to the following equations:
\begin{eqnarray}
d\omega_{1}   =0\\
d\omega_{2n}  =\sum_{k=2}^{n}\left(  -1\right)  ^{k}\omega_{k}\wedge
\omega_{2n+1-k}+2\lambda_{2}\omega_{2n}\wedge\theta
\end{eqnarray}
Then the sum $\omega=\omega_{1}+\omega_{2n}$ satisfies
\begin{equation}
\omega\wedge\left(  \bigwedge^{n}d\omega\right)  =2n!\omega_{1}\wedge
...\wedge\omega_{2n}\wedge\theta\neq0,
\end{equation}
and defines a contact form.
\end{enumerate}
\end{proof}

In \cite{Ree} it was shown that contact forms $\alpha$ on varieties imply the existence of a vector field $X$ such that $\alpha(X)=1$ and $X\lrcorner \alpha=0$, called the dynamical system associated to $\alpha$. Therefore for the previous solvable Lie algebras we can construct dynamical systems, which moreover have no singularities \cite{Ree}. On the contrary,  solvable Lie algebras having the nilpotent graded algebra $\frak{n}_{n,1}$ as nilradical have a number of invariants which depends on the dimension, which implies that (in odd dimension) they cannot possess a contact form \cite{C24}. This loss of structure is due to the fact that the Heisenberg subalgebra of $Q_{2n}$ spanned by $\left\{X_{2},..,X_{2n}\right\}$ is contracted onto the maximal abelian subalgebra of $\frak{n}_{n,1}$.

\section{Quasi-classical subalgebras}

In \cite{Ok1} the notion of quasi-classical Lie algebras was introduced to
present abelian and semisimple gauge theories in a unified manner. Moreover, this approach allows to
construct gauge theories based on non-abelian and non-semisimple Lie algebras. Inspite of the objection of having ghosts when the compacity condition \footnote{That is, when the Lie algebra is reductive.} is not satisfied, quasi-classical algebras are still of interest for integrable models with the nonzero curvature condition, as well as for some solutions of the Yang-Baxter equations \cite{Ok1,Das}. In this section we show that the nilpotent graded Lie algebras $Q_{2n}$ analyzed in this paper, as well as the solvable Lie algebras with $Q_{2n}$-nilradical always contain a maximal non-abelian quasiclassical Lie algebra of dimension $2n-1$.

\medskip

A Lie algebra $\frak{g}$ is called quasi-classical if it has a bilinear, associative,
symmetric and non-degenerate form $H\left(  \;,\right)  $. Obviously any
reductive Lie algebra satisfies the requirement, and is therefore
quasi-classical. However, non-reductive algebras of this type exist. In
\cite{Ok1} a characterization in terms of quadratic Casimir operators was given:

\begin{proposition}
A Lie algebra $\frak{g}$ is quasi-classical if and only if it possesses a
quadratic Casimir operator $I_{2}=g^{\alpha\beta}X_{\alpha}X_{\beta}$ such
that the symmetric matrix $\left(  g^{\alpha\beta}\right)  $ has an inverse
$\left(  g_{\alpha\beta}\right)  $ satisfying
\begin{equation*}
g^{\alpha\beta}g_{\alpha\beta}=\delta_{\alpha\beta}.
\end{equation*}
\end{proposition}

It follows in particular that the invariants of a quasi-classical Lie algebra
depend on all its generators.

\begin{proposition}
For any $n\geq3$, the nilpotent algebra Q$_{2n}$ contains a maximal
nonabelian  quasi-classical subalgebra of dimension $\left(  2n-1\right)  $.
\end{proposition}

\begin{proof}
From proposition $2$ we know that for any value of $n$, the nilpotent algebra
$Q_{2n}$ has the quadratic invariant
\begin{equation}
I_{2}=x_{1}x_{2n}+x_{3}x_{2n-1}+\sum_{j=4}^{n}\left(  -1\right)  ^{j}%
x_{j}x_{2n+2-j}+\frac{\left(  -1\right)  ^{n+1}}{2}x_{n+1}^{2}.
\end{equation}
This function actually coincides with its symmetrization, since the involved
variables correspond to commuting generators of the algebra. Therefore we can write the Casimir
operator in matrix form:
\begin{equation}
\fl I_{2}=\left(  X_{1},X_{2},..,X_{2n-1},X_{2n}\right)  \left(
\begin{array}
[c]{cccccccccc}%
&  &  &  &  &  &  &  &  & 1\\
&  &  &  &  &  &  &  & 0 & 0\\
&  &  &  &  &  &  &  & 1 & \\
&  &  &  &  &  &  & -1 &  & \\
&  &  &  &  &  & . &  &  & \\
&  &  &  &  & \frac{\left(  -1\right)  ^{n}}{2} &  &  &  & \\
&  &  &  & . &  &  &  &  & \\
&  &  & -1 &  &  &  &  &  & \\
& 0 & 1 &  &  &  &  &  &  & \\
1 & 0 &  &  &  &  &  &  &  &
\end{array}
\right)  \left(
\begin{array}
[c]{c}%
X_{1}\\
X_{2}\\
:\\
\\
\\
\\
:\\
X_{2n-1}\\
X_{2n}%
\end{array}
\right).
\end{equation}
The matrix is obviously symmetric, but of rank $2n-1$. However, and since the
invariants of $Q_{2n}$ do not depend on the generator $X_{2}$, we can consider
the subalgebra $\frak{k}_{n}$ of $Q_{2n}$ generated by $\left\{  X_{1}%
,X_{3},..,X_{2n}\right\}  $. From the system (\ref{eq1})-(\ref{eq3}) it follows
at once that any invariant of $Q_{2n}$ is also an invariant of $\frak{k}_{n}$,
and since the centre of this subalgebra has dimension 2, it has the
supplementary invariant $x_{2n-1}$. In particular, the quadratic Casimir
operator $I_{2}$ of $\frak{k}_{n}$ can be written as
\begin{equation}
\fl I_{2}=\left(  X_{1},X_{3},..,X_{2n-1},X_{2n}\right)  \left(
\begin{array}
[c]{ccccccccc}%
&  &  &  &  &  &  &  & 1\\
&  &  &  &  &  &  & 1 & \\
&  &  &  &  &  & -1 &  & \\
&  &  &  &  & . &  &  & \\
&  &  &  & \frac{\left(  -1\right)  ^{n}}{2} &  &  &  & \\
&  &  & . &  &  &  &  & \\
&  & -1 &  &  &  &  &  & \\
& 1 &  &  &  &  &  &  & \\
1 &  &  &  &  &  &  &  &
\end{array}
\right)  \left(
\begin{array}
[c]{c}%
X_{1}\\
X_{3}\\
:\\
\\
\\
\\
:\\
X_{2n-1}\\
X_{2n}%
\end{array}
\right)
\end{equation}
over the basis $\left\{  X_{1},X_{3},..,X_{2n}\right\}  $,
showing that this algebra is quasi-classical.
\end{proof}

\begin{remark}
For the Lie algebra $\frak{n}_{n,1}$ we also find a quasi-classical maximal
subalgebra, and, as follows from the structure of their invariants \cite{Wi},
this subalgebra is necessarily abelian. This means that for the corresponding quasi-classical subalgebras, the contraction of $Q_{2n}$ onto $\frak{n}_{2n,1}$ recovers the abelian gauge theory.
\end{remark}

\begin{corollary}
Any solvable Lie algebra with nilradical isomorphic to $Q_{2n}$ possesses a nonabelian quasi-classical Lie algebra of dimension $2n-1$.
\end{corollary}

In fact it follows from the structure of these algebras, that none of them is quasiclassical, since only $\frak{r}_{2n+1}(1-n)$ has a quadratic Casimir operator. However, this does not define a non-degenerate form since the invariant is independent on the variable associated to the torus generator (or has no invariant if the maximal torus is added).

\section{Conclusions}

We have completed the study of the generalized Casimir invariants of indecomposable solvable real Lie algebras with a   naturally graded nilradical of maximal nilindex initiated in \cite{Wi}. Although $Q_{2n}$ is a contraction of $\frak{n}_{n,1}$, the corresponding solvable Lie algebras obtained exhibit rather different structural properties. In particular, there is no relation by contraction between these algebras, up to the case where both nilpotent algebras have the same torus of derivations. That is, only $\frak{r}_{2n+1}(1)$ contracts onto a solvable Lie algebra with $\frak{n}_{n,1}$-nilradical. Further, while the number of invariants of the algebra $\frak{n}_{n,1}$ depends on the dimension, for $Q_{2n}$ it remains fixed for any dimension, and coincides with the maximal number of nil-independent derivations. As a consequence, the corresponding solvable Lie algebras have only one invariant, which for special values reduces to a classical Casimir operator, or none invariants if both nil-independent elements are added.  This fact implies the existence of a contact form on the corresponding solvable Lie algebras or rank one, and is of potential interest in connection with their contractions onto the Heisenberg Lie algebra \cite{C24} and the construction of positive Einstein metrics \cite{Bo}. As expected, most of the invariants of solvable algebras with $Q_{2n}$-nilradical are harmonics, and for some nondiagonal derivations logarithmic functions appear.

Another interesting fact is that $Q_{2n}$ and the associated solvable algebras contain a maximal nonabelian quasi-classical Lie algebra of codimension one, respectively two. This follows from the structure of the quadratic Casimir operator of the nilradical, and is strongly related to the maximal Heisenberg subalgebra of $Q_{2n}$. The latter subalgebra constitutes the reason for the main difference between solvable Lie algebras with nilradical $\frak{n}_{n,1}$ and $Q_{2n}$ when realized as symmetry algebras of differential equations. As known, the Heisenberg algebra $\frak{h}_{n}$ of dimension $2n+1$ can be faithfully realized in $k\geq (n+1)$-dimensional space by vector fields. Therefore any Lie algebra that contains $\frak{h}_{n}$ will need at least $(n+1)$-variables for any realization by vector fields. Since $Q_{2n}$ contains $\frak{h}_{n-1}$ for any $n\geq 3$, the solvable Lie algebras with $Q_{2n}$-nilradical will appear as symmetries of partial differential equations in $k$-dimensional space, where $k\geq n$. As the algebra $Q_{2n}$ is defined for $n\geq 3$, these solvable algebras do not describe dynamics of physical systems given by a system of ordinary differential equations. On the contrary, solvable Lie algebras with $\frak{n}_{n,1}$-nilradical allow planar realizations, and can therefore appear as symmetries of ordinary differential equations. Due to the simplicity of the invariants and structure for both cases, solvable Lie algebras with naturally graded nilradical of maximal nilindex are suitable candidates to analyze the problem of superposition formulae for nonlinear differential equations \cite{Wi3}.

Finally, solvable Lie algebras analyzed in this article and in \cite{Wi} are also of interest in the reduction of sourceless Yang-Mills equations by means of potentials with constant components. Since the nilpotent algebra $\frak{n}_{n,1}$ contains an abelian ideal of codimension one \cite{Wi}, any rank one solvable Lie algebra associated to this nilradical will contain a codimension one solvable Lie subalgebra that only admits flat Yang-Mills potentials. By contrast, no solvable Lie subalgebra of solvable Lie algebras having $Q_{2n}$ as nilradical has this property. Indeed, since these algebras always contain a Heisenberg subalgebra of dimension $2n-1$, by a result of \cite{Mun}, any Lie algebra containing it necessarily admits a nonflat Yang-Mills potential. In conclusion, physically the solvable Lie algebras of proposition 5 exhibit a different behaviour from those of paper \cite{Wi}. This fact moreover suggests that, even in the solvable case, group theoretical arguments based on graded Lie algebras are an adequate frame to analyze models for different physical phenomena.

\section*{Acknowledgements}
The authors wish to express their gratitude to P. Winternitz for
numerous fruitful discussions and the interest to the subject, as
well as for several improvements of the manuscript.\newline During
the preparation of this work, the authors were partially supported
by the research project PR1/05-13283 of the U.C.M.


\section*{References}

\begin{thebibliography}{99}

\bibitem{Ba} Barut A O and Raczka R 1980 \textit{The theory of group representations and applications} (Warsaw: PWN Polish Scientific publishers)

\bibitem{Lya} Lyakhovskii' V D and Bolokhov A A 2002 \textit{Gruppy simmetrii i elementarnye chastitsy} (Moscow: URSS)

\bibitem{Pet} Petrov A Z 1969 \textit{Einstein spaces} (Oxford: Pergamon)

\bibitem{Sm} Schmutzer E 1972 \textit{Symmetrien und Erhaltungss\"atze der Physik} (Berlin: Akademie-Verlag)

\bibitem{Ar} Arkhangel'skii' A A 1979 \textit{Mat. Sb.} \textbf{108} 134

\bibitem{Ok2} Okubo S 1998 \JPA \textbf{31} 7603

\bibitem{Mo} Morozov V V 1958 \textit{Izv. Vys. Uchebn. Zav. Mat.} \textbf{5} 161

\bibitem{Mu63} Mubarakzyanov G M 1963 \textit{Izv. Vys. Uchebn. Zav. Mat.} \textbf{32} 114

\bibitem{Tu}  Turkowski P 1988 \JMP \textbf{29} 2139

\bibitem{Wi1} Ndogmo J and Winternitz P 1994 \JPA \textbf{27} 2787

\bibitem{Wi2} Tremblay S and  Winternitz P 2001 \JPA \textbf{34} 9085

\bibitem{Ca3} Campoamor-Stursberg R 2003 \JMP \textbf{44} 771

\bibitem{Wi} \v{S}nobl L, Winternitz P 2005 \JPA \textbf{38} 2187

\bibitem{Ok3} Okubo S and Kamiya N 2002 \textit{Comm. Algebra} \textbf{30} 3825

\bibitem{Tro} Trofimov V V 1979 \textit{Izv. Akad. Nauk SSSR, Ser. Mat.} \textbf{43} 714

\bibitem{Be}  Beltrametti E G and Blasi A 1966 \textit{Phys. Lett.} \textbf{20} 62

\bibitem{Pe} Pecina J N 1994 \JMP \textbf{35} 3146

\bibitem{C43}  Campoamor-Stursberg R 2004 \textit{Phys. Lett. A} \textbf{327} 138

\bibitem{Sh} Patera J, Sharp R T, Winternitz P and Zassenhaus H 1976 \JMP \textbf{17} 986

\bibitem{Vr} Vergne N 1970 \textit{Bull. Soc. Math. France} \textbf{98} 81

\bibitem{C48} Campoamor-Stursberg R 2005 \textit{Alg. Colloqiuim} \textbf{12} 497

\bibitem{Kru} Kruglikov B 1998 \textit{Proc. Steklov Math. Inst.} \textbf{221} 232

\bibitem{Ree} Reeb G 1952 \textit{Mem. Acad. Sci. Bruxelles} \textbf{27} 1

\bibitem{C24} Campoamor-Stursberg R 2003 \textit{Acta Phys. Pol. B} \textbf{34} 3901

\bibitem{Ok1} Okubo S 1979 \textit{Hadronic J.} \textbf{3} 1 

\bibitem{Das} Das A 1989 \textit{Integrable models} (Singapur: World Scientific)

\bibitem{Bo} Boyer C P and Galicki K 2000 \textit{Int. J. Math.} \textbf{11} 873

\bibitem{Wi3} Winternitz P 1993 Lie groups and solutions of nonlinear partial differential equations \textit{integrable systems, Quantum groups and quantum field theories}  (Dordrecht: Kluwer), pp 515-567

\bibitem{Mun} Schimming R and Mundt E 1992 \JMP \textbf{33} 4250

\end{thebibliography}
\end{document}